\documentstyle[aps,epsfig,amstex]{revtex}

\begin{document}

\title{Quantification of Sleep Fragmentation Through the Analysis of
Sleep-Stage Transitions}

\author{Chung-Chuan Lo$^{1}$, Plamen Ch. Ivanov$^{1,2}$,
Lu\'{\i}s A. Nunes Amaral$^{1,2}$, \\
Thomas Penzel$^{3}$, Claus F. Vogelmeier$^{3}$ and H. Eugene
Stanley$^{1}$
}

\address{$^1$Center for Polymer Studies and Department of Physics
           Boston University, \\Boston, MA 02215, USA \\
     $^2$Cardiovascular Division, Beth Israel Deaconess Medical
       Center, \\
       Harvard Medical School, Boston, MA 02215, USA \\
     $^3$Klinik f\"ur Innere Medizin - Pneumologie, Philipps-Universit\"at
       Baldingerstrasse 1, Marburg D-35033, Germany \\ }


\maketitle


\begin{center}
{\bf ABSTRACT}
\end{center}
{\noindent {\bf Study Objectives:} We introduce new quantitative
 approaches to study sleep-stage transitions with the goal of
addressing the two following questions: (i) Can the new approaches
provide more information on the structure of sleep-stage
transitions? (ii) How does sleep fragmentation in patients with
sleep apnea affect the structure of sleep-stage transitions?
\\ {\bf Design:}
We analyze hypnograms and compare normal subjects and sleep apnea
patients using numerous measures, including the percentage of sleep
time for each stage, probability distributions of the duration of
each stage, the sleep-stage transition matrix, and a measure of the
asymmetry of this matrix. \\ 
{\bf Setting:} N/A \\ 
{\bf Subjects:}
197 normal subjects and 50 obstructive sleep apnea patients
recruited in the SIESTA project.\\
{\bf Results:} We find that the time percentage for wake stage is
identical for sleep apnea subjects and for normal subjects, but
that the sleep apnea group have a faster decaying distribution of wake
duration. Both normal subjects and sleep apnea patients have
exponential distributions of duration for all sleep stages and a
power law for the wake stage. We also find that there is a loss of
preference of transition paths of sleep stages in sleep apnea.
\\
{\bf Conclusions:}
The new approaches proposed here enable us to show that the distribution
of sleep and wake duration have different functional forms, indicating
fundamental differences in the dynamics between sleep
and wake control. The difference remains even in the fragmented sleep
of sleep apnea. The fragmentation of sleep in sleep apnea results
in a shorter wake duration and interrupts the structure of
sleep-stage transitions of sleep apnea subjects, causing the loss of certain
particular transition paths.
\\
}


\section*{introduction}

Analyses of sleep-stage transitions have long been used as
diagnostic tools in
clinical applications. Such analyses mostly
concentrated on the changes in the time percentage for
each sleep stage and in other simple statistics such as the total
number of arousals during nocturnal sleep
~\cite{Chokroverty-S-1999a,Carskadon-M-2000a}.
There have also been several studies focusing on statistical measures
such as transition probabilities
~\cite{Williams-R-1964a,Brezinova-V-1975a,Kemp-B-1986a,Yassouridis-A-1999a,Karlsson-M-2000a},
but many other statistical properties of sleep-stage
transitions have not been considered in a systematic way.

Sleep-stage transitions are sometimes described
as having quasi-cyclic behavior (the sleep cycle)
~\cite{Carskadon-M-2000a},
but on top of the periodic patterns, there are many transitions 
without apparent periodicity (Fig. ~\ref{f.hypnograms}).
Indeed, even if one disregards
all sleep-stage transitions and considers only the
wake stage during sleep, one still finds intriguing statistical
properties and no apparent periodicity ~\cite{Lo-C-2002a}.
Furthermore, it has been reported that
sleep stages correlate with the dynamics of the autonomic nervous
system. For example, the correlations and scaling behavior in
heart-rate variability depends on
sleep stages ~\cite{Ivanov-P-1999b,Bunde-A-2000a}.
Because sleep-stage transitions are such complex
processes, simple statistical measures may not be sufficient to
describe their dynamics and uncover any information contained
in the fluctuations. Therefore, we study
sleep-stage transitions with methods from modern statistical
physics and nonlinear dynamics.

Many advanced statistical analyses have been applied to
the study of the electroencephalogram (EEG) during sleep
~\cite{Fell-J-1996b,Pradhan-N-1996a,Pereda-E-1998a,Fell-J-2000a},
but an important limitation of these methods is that
the EEG records only the activity close to the
cortex surface, while it is believed that sleep is regulated
by neurons in the hypothalamus \cite{Saper-C-2001a}.
Hence, we hypothesize that to study the dynamics
of sleep regulation, one must investigate sleep-stage
transitions, which contain more global information,
including not only the EEG, but also eye movements and
muscle tone.

There are two major limitations in the analysis of
sleep-stage transitions: The first is the limited number of data points
($\approx 900$ points per night, where each point represents the
sleep stage in a epoch of 30 seconds).  The second is the
discretization of the data into six sleep stages. These limitations constrict
the mathematical
tools which can be used in the analysis of sleep-stage transitions,
so we focus on the distributions of duration of sleep stages, the transition
probability matrices, and the degree of asymmetry of these matrices.

We will also address questions regarding the statistical properties
that we find: (i) how do these statistical properties change under the
influence of sleep disorders, and (ii) which of these properties are 
fundamental and do not change under the influence of sleep disorders?
To this end, we also study subjects with obstructive sleep apnea, who 
experience
fragmented sleep with a reduced amount of slow-wave sleep and more
awakenings (Fig. ~\ref{f.hypnograms}c). The sleep fragmentation is
characterized by large number of arousals during
nocturnal sleep. When arousal periods are longer than 15 seconds within a
30-second epoch of observation, the epoch is classified as a wake stage.
The fragmentation of sleep in obstructive sleep apnea arises from
respiratory problems
~\cite{Strollo-P-1996a,Robinson-A-1999a}.
Therefore, sleep apnea is a good model for studying the effect of
external disturbances on sleep-stage transitions.

In the present study, we propose new quantitative approaches to studying
sleep-stage transitions. We show that these approaches enable us
to find more information on the structure of sleep-stage transitions
and enable us to find how sleep fragmentation
of sleep apnea affects the structure of the sleep-stage transitions.
Thus, the present approach gives us additional insights into the dynamics
of sleep and wakefulness.


\section*{Methods}
\subsection{Subjects and Data acquisition}
{We analyze a database comprising 197 normal subjects and 50
patients with obstructive sleep apnea collected in eight leading
European sleep laboratories under the SIESTA project
\cite{Klosch-G-2001a}. For each subject, two consecutive nights
were recorded with cardiorespiratory polysomnography. Sleep stages
were determined according to the Rechtschaffen and Kales criteria
\cite{Rechtschaffen-A-1968a}: two channels of electroencephalogram
(EEG), two channels of electrooculogram (EOG), and one channel of
submental electromyogram (EMG) were recorded. Signals were
digitized at a minimum of 100 Hz, and a 12-bit resolution, and are
scored visually in epochs of 30 seconds for six stages:
wakefulness, rapid-eye-movement (REM) sleep, and
non-rapid-eye-movement (NREM) sleep stages 1, 2, 3, and 4.
Subjects wend to bed at midnight and were allowed to
wake up in the morning at their own well. The average sleep
time are xxx for healthy subjects and xxxx for sleep apnea
subjects. Wake periods prior to the first sleep stage and
after the last sleep stage are excluded from the analyses.

We analyze hypnograms of the second night only. In order to
eliminate the effect of age on sleep, we choose 48 of 197 normal
subjects and 48 of 50 sleep apnea subjects. The reason for removing
two sleep apnea subjects from the group is that these two sleep
apnea subjects were much older (74) or younger (29) than the other
sleep apnea subjects. We select the normal subjects according to
the following procedure: We choose normal subjects from sleep
laboratories which also provide sleep apnea subjects. The subjects
are chosen to match the ages of 48 sleep apnea subjects. After all
age-matched subjects have been chosen, we
choose normal subjects randomly from other laboratories, also with
similar age, until 48 normal subjects have been selected.

We are thus able to choose normal and sleep apnea subjects with
matched ages and maximum possible overlap of source laboratories.
The selected normal group has an average age of $50.9$ and a standard
deviation of $9.4$, while the selected sleep apnea group has an
average age of $51.3$ and a standard deviation of $8.9$. We use the
entire database of normal subjects in a test of the reliability of
our results.

\subsection{Coarse-graining of sleep stages}

A major difficulty of studying the statistical properties
of sleep-stage transitions is inter-scorer reliability,
a topic of great concern in the literature
\cite{Kelley-J-1985a,Kubicki-S-1989a,Whitney-C-1998a,Norman-R-2000a,Kunz-D-2000a}.
One study reports that the agreement between
sleep-stage scorers are in the range of 30\%--90\%, depending on
the sleep stages and ages of subjects \cite{Kunz-D-2000a}.
The least reliable scoring occurs for the NREM
stage 1, which has only a 38\% agreement on average. All other stages,
such as wake, slow-wave sleep, and REM sleep, have
an average agreement higher than  70\%. To minimize scoring uncertainty,
we reduce the six scored stages of sleep
into four stages:  We keep wake and REM stages unchanged, and group stages
1 and 2 into a single stage (light sleep),
and stages 3 and 4 into a single stage (slow-wave sleep). The stages,
wake, light sleep, slow-wave sleep, and REM sleep are abbreviated
as W, L, S, and R, respectively.

\subsection{Percentage of sleep time}
We define $F_m$ to the percentage of total sleep time for
sleep stage $m$.  We measure $F_m$ for each sleep stage for each
subject, and then calculate the mean and standard deviation
of $F_m$ for normal and sleep apnea groups. We apply
Student's t-test to determine the level of significance of
the differences in $F_m$ between normal and sleep apnea groups.

\subsection{Distributions of duration}
$F_m$ is a useful tool in diagnostic application, but
it cannot capture all the information about the sleep-stage
transitions. For example, identical values of $F_m$ could result from
many short periods or from just a few long periods; two situations
with different underlying dynamics.
Therefore, we study the distributions of the duration of wake and sleep
stages for normal and sleep apnea groups.
The distribution of duration of events is a
useful measure for studying the underlying dynamics of a system.
For example, a peak-like distribution indicates
the periodic occurrence of events with fluctuations,
such as heart beats intervals \cite{Peng-C-1993a,Ivanov-P-1996a}.
An exponential distribution
is usually associated with a random process, such as
fluorescent decay \cite{Campbell-I-1984a}. A power-law
distribution, which has been observed in many systems, such as
earthquakes \cite{Bak-P-2002a}, solar flares \cite{Boffetta-G-1999a},
and rainfall \cite{Peters-O-2002a},
is associated with self-organized criticality
\cite{Bak-P-1996a,Sanchez-R-2002a} or other complex
mechanisms ~\cite{Sornette-D-2000a}.
Thus, in order to quantitatively study how the brain
regulates sleep, we consider the distribution of duration of sleep
stages.

We first calculate the duration of the separate wake and sleep stages
periods for each subject, then pool the data from all subjects and
calculate the group's cumulative distributions of duration for
wake, light sleep, slow-wave sleep, and REM sleep, which we denote
as $P_W(d)$, $P_L(d)$, $P_S(d)$, and $P_R(d)$, respectively
(Appendix A).

\subsection{Transition probability matrices}
The percentage $F_m$  and the distribution
of duration $P_m(d)$ of each sleep stage
provide important information
about the sleep stages, but they cannot provide temporal
information -- i.e., time organization of transitions.
For example, $P_m(d)$ does not reveal any information
about the preferred path of the transitions.
Hence, we must study the transition probabilities between
different sleep stages.

Several types of transition probabilities for sleep stages
have been studied
~\cite{Kemp-B-1986a,Yassouridis-A-1999a,Karlsson-M-2000a}.
Here, we consider a new type of transition probability
matrix $T$ with elements $T_{mn}$, defined as $T_{mn}=N_{mn}/N$,
where $N_{mn}$
is the number of transitions from stage $n$ to stage $m$
during the entire night and $N$ is the total number of
transitions. We calculate the matrix $T$ for each
subject and then calculate the means and the standard errors
of $T_{mn}$ for normal and sleep apnea groups.

The matrix $T$ is particularly useful for the analysis
of the local structure of transitions because it measures the
transition probability between two consecutive stages.
Since there are a large number of short transitions
between different sleep stages even for normal subjects
(Fig. \ref{f.hypnograms}), the
transition matrix may be able to reveal hidden patterns in
these short transitions.

\subsection{The coefficient of asymmetry of transitions}
One of the advantages of studying transition probabilities
is that one may be able to extract the information about
locally preferred paths in the sleep-stage
transitions. One important question is: If there are
locally preferred paths, are they affected
by the sleep fragmentation of the sleep apnea?
To answer this question, we introduce the
concept of the symmetry of transitions. When the probability
of having a transition from state A to state B is equal to
the probability of having one from B to A, transitions between A and B
are called ``symmetric''. On the contrary, if the probability
from A to B is not equal to the probability from B to A, the
transition is called ``asymmetric'', a pattern indicating
a preferred local transition path. In order to quantify
the asymmetry in the sleep-stage transitions, we introduce
the coefficient of asymmetry $A$ (Appendix B), which can be
calculated from the transition matrix. $A=0$
corresponds to a completely symmetric matrix, while
$A=1$ corresponds to a completely asymmetric matrix.
We calculate $A$ for all subjects and perform Student's t-test
to evaluate the significance of the measured differences in
$A$ for normal and sleep-apnea groups.

\subsection{Test of the reliability of the results}
Since sleep stages are scored visually by different experts in
each laboratory, it is important to evaluate how much human
bias affects the statistical analyses carried out. We compare our
results with the data provided by different laboratories. For
example, for the fraction $F_m$ of total sleep time for a given
sleep stage $m$, we calculated $F_m$ for each subject, and compare
the distribution of $F_m$ of normal subjects from one laboratory
to the normal subjects from a different laboratory. We also
compare $F_m$ of normal subjects from one laboratory to the sleep
apnea subjects from the same laboratory and from a different
laboratory. We perform the same procedure on all other statistical
measures calculated in this study. We calculate the $p$ value
for Student's t-test \cite{Press-W-1994a} and find the differences
between normal and apnea subjects are much more significant ($p<0.05$)
than the differences between normal subjects from different
laboratories ($p>0.05$).


\section*{Results}

We first show results for the time percentage $F_m$ for each sleep
stage for both normal and sleep apnea groups
(Fig.~\ref{f.fraction}). We find significant differences between
the two groups for stage 1 ($p<0.001$) and also for stage 4
($p<0.05$). However, there are no significant differences between
normal and sleep apnea groups for the wake stage, sleep stage 2, and
REM sleep.

In Fig.~\ref{f.distdur} we show distributions of duration
for wake, $P_W(d)$, light sleep, $P_L(d)$, slow-wave sleep,
$P_S(d)$, and REM sleep, $P_R(d)$, for normal
and sleep apnea groups. $P_W(d)$ shows a power-law decay
$P_W(d)\propto d^{-\alpha}$ for both normal and sleep apnea
groups, while $P_L(d)$, $P_S(d)$, and $P_R(d)$ all show exponential
decays, $P(d)\propto e^{-d/\tau}$.

We compare $P_W(d)$, $P_L(d)$, $P_S(d)$, and $P_R(d)$ for normal
and sleep apnea subjects and find that (i) $P_W(d)$ for sleep apnea
group has a larger exponent $\alpha=1.28\pm 0.03$ than the normal group,
$\alpha=1.11\pm 0.05$, and (ii) $P_L(d)$, $P_S(d)$, and $P_R(d)$ for the
sleep apnea group show a steeper decay in the region $d<5$ min,
but their decay time scale $\tau$ in the
longer time region are similar to those of the normal group.

We show the transition probability matrices $T$
for the normal and sleep apnea groups in Tables 1a\& b.
Not surprisingly, for both normal and sleep apnea subjects, we find
some transitions are virtually prohibited: (i) there are no
$R \rightarrow S$ transitions and only few $S \rightarrow R$ transitions,
and (ii) there is no $W \rightarrow S$ transition.

For both the normal and sleep apnea groups, the matrices are asymmetric,
but this asymmetry decreases for the sleep
apnea group. To quantify the degree of asymmetry for normal and sleep
apnea groups, we calculate the coefficient of asymmetry $A$ for the 
data from each subject and plot distributions of $A$
for normal and sleep apnea groups (Fig.~\ref{f.asymmetry}).
The normal group has a mean $A$ of $0.058 \pm 0.004$, while the sleep
apnea group has a mean $A$ of $0.034 \pm 0.003$.

We perform Student's t-test to calculate the level of
significance of the difference in $A$ between normal and sleep
apnea groups, resulting in a $p$ value smaller than $1\times 10^{-5}$.

We also test to see if $A$ changes for
elder subjects, who are known to show more arousals during
nocturnal sleep ~\cite{Chokroverty-S-1999a}.
We choose two groups from our database of 197 normal subjects:
young (47 subjects, age: 20--35 years), and old (52 subjects age: 60--75 years).
We find that the number of sleep-stage transitions in the older group has
a mean of $106\pm3$, which is significantly larger than the mean of
$84\pm2$ for the young group.

We compare distributions of $A$ between these
two groups. The young group has an average $A$ of $0.061\pm 0.005$, while
the old group has an average $A$ of $0.051\pm 0.004$. Applying Student's
t-test, we find $p=0.6$.

A comparison of $F_m$, $P_m(d)$ and $A$ of normal group
with sleep apnea group is shown in Table 2.


\section*{Discussion}

We have proposed new approaches to the characterization
of the dynamics of sleep-stage transitions, and found several
intriguing properties:

\begin{enumerate}
\item {
We find that for normal subjects, the duration of each sleep stage
is characterized by an exponential distribution with specific time
scale, and the duration of wake stage is characterized by a
power-low distribution suggesting a scale-free dynamics. This
finding suggests a fundamental difference between the dynamics of
sleep and wakefulness control. It implies that sleep and
wakefulness are not just two parts of a sleep-wake control, but
that there exist entirely different mechanisms for their
regulation in the brain, which supports recent studies in the
neuronal level of sleep mechanisms (see, e.g., Ref.
\cite{Saper-C-2001a}).

Ref. \cite{Lo-C-2002a} reported that, for normal subjects, the duration
of wake periods follows a  power-law distribution,
$P_{W}(d)\propto d^{-\alpha}$,
while the duration of sleep periods follows an exponential distribution,
$P(d)\propto e^{-d/\tau}$.
As shown in Fig.~\ref{f.distdur}, when
we decompose sleep into three stages: light, slow-wave and REM sleeps, all
these sleep stages still follow exponential distributions.
It is surprising that REM sleep, which can be regarded as somehow
similar to wake from a brain activity aspect, clearly follows an
exponential distribution of duration as the rest of the sleep
stages, which is different from the power-law distribution of
duration of the wake stage.

The same forms of the distributions are observed for sleep apnea
patients for all sleep stages (exponential) and for the wake stage
(power law). This finding suggests robust mechanisms of sleep and
wakefulness controls which do not change with sleep fragmentation
in sleep apnea.}

\item{Our finding that the time percentage $F_W$ for the wake stage of the
sleep apnea group is not significantly different
from that of the normal group appears to contradict the ``common''
expectation that sleep apnea subjects have more
arousals. However, the difference in the wake stage between normal
and sleep apnea subjects is clearly observed in the distributions
of wake duration $P_W(d)$.
The difference in the values of the power-law exponent $\alpha $
characterizing $P_W(d)$ (Fig.~\ref{f.distdur}) indicates that
wake periods for sleep apnea subjects have
shorter duration. Since $F_W$ is identical, sleep apnea subjects
must have a larger number of wake periods.
This is a clear indication of the sleep fragmentation one expects
for sleep apnea.

Although the functional form of $P_L(d)$, $P_S(d)$ and $P_R(d)$
is identical for normal and sleep apnea groups, the characteristic
time scales are different, except for REM sleep. We find that the
most significant change
occurs for short duration (Fig.~\ref{f.distdur}b\&d).
The increasing of slopes in the short duration $d<5$ min,
indicates that sleep apnea subjects have many more short
stages than normal subjects do, and thus a more fragmented sleep.

Note that the power-law exponent $\alpha=1.1$ for $P_W(d)$ for normal
subjects (Fig.~\ref{f.distdur}b).
This value is different from what we reported ($\alpha=1.3$) previously
~\cite{Lo-C-2002a}. The reason is that in Ref. \cite{Lo-C-2002a}
our results was based on the database of 20 young subjects with average
age of $35.2$, which is different from the
average age of $ 50.9$
of the 48 normal subjects we used in this study. With the choice of
young normal subjects from the database we used in the present study,
we recover $\alpha=1.3$, which in agreement with the value
reported in Ref. \cite{Lo-C-2002a}.}

\item{From the transition matrix $T$ we find that the transition
probabilities between several pairs of stages change with sleep
apnea. These changes can be characterized by the coefficient
of asymmetry. Both normal and sleep apnea groups have
asymmetric transition matrix $T$, but the sleep apnea group
exhibits an increase of symmetry.
The implication of an asymmetric
transition matrix is that the transition process has
preferred transition paths. Comparing
$T_{RL}$ and $T_{LR}$, we find that there are more transitions
from light
sleep to REM sleep than from REM sleep to light sleep. We also find,
by comparing $T_{WR}$ and $T_{RW}$, that there are more transitions
from REM to wake then from wake to REM. These findings indicate that
when a $R\rightarrow W$ transition occurs, the sleep control system
``prefers'' to make a transition to light sleep instead of back to
REM (Fig.~\ref{f.illust}a).
The explanation is supported by the values of $T_{LW}$ and $T_{WL}$:
there are more $W\rightarrow L$ transitions than
$L \rightarrow W$ transitions.

We find that the matrices exhibit increased symmetry
for the sleep apnea group (Fig.~\ref{f.asymmetry}). This indicates
that the sleep-stage transitions of sleep apnea subjects have less
local structure. From the distributions
of wake duration, we learn that sleep apnea subjects
have a larger number of transitions but shorter duration. The increased
wake periods,
according to transition matrices, increase the symmetry of
the matrix by distributing with less preference throughout the night
(Fig.~\ref{f.illust}b).}

\item{A question one may ask is if the increase of the symmetry
is a necessary result of the increase of the number of wake periods?
As described in the
results section, we calculate $A$ for elderly subjects which have
significantly larger number of wake periods during sleep.
It is very interesting that although elderly subjects experience a
larger number of wake periods, the coefficient of asymmetry
does not change significantly ($p=0.06$). This
might indicate that the preferred transition path
observed in normal subjects is fundamental, and is not significantly
affected by age: The increased wake periods do not significantly
change the preference of sleep-stage transition in elder subjects,
while the increased wake periods in sleep apnea subjects do.}

\end{enumerate}

All of our analyses are based on group distributions.
However, both normal and sleep apnea groups have broad
distributions for many statistical measures.
It is not known whether the changes in group distributions
are representative of changes in the individual behavior. It is
also not known if each individual in the normal group (or in the
sleep apnea group) follows the same statistics.
To answer the questions, data of at least ten
nights from each subject are needed. One can then compare the
distribution of statistical measures from data on one subject
to the data on another subject.

Furthermore, all the analyses are based on whole-night records.
However, sleep is not a homogeneous process. The
statistical properties may vary throughout the
night ~\cite{Born-J-1999a,Carskadon-M-2000a,Lo-C-2002a}. Hence,
it is important to study the changes in $P_m(d)$, $T$ and
$A$ in the course of the night.

Our findings of the stability of underlying dynamics
of sleep-stage transitions between normal and sleep
apnea subjects are intriguing. It is important to test
if the dynamics changes under pharmacological influences
such as sleep-inducing drugs or caffeine, or under different
psycho-physiological or pathological conditions such as
stress or depression.

\newpage


\section*{Appendix}

\subsection*{A. Cumulative distribution of duration}

Let $p_m(d)$ be the probability density function (i.e. the
probability distribution) for the duration $d$ of a given stage $m$
for the group. We study the cumulative distribution $P_m(d)$,
which is defined as:

\begin{equation*}
P_m(d) \equiv \int_d^\infty p_m(r) dr \,.
\label{eq.cumulative}
\end{equation*}

{\noindent Therefore, $P_m(d)$ is the probability of having a period of
stage $m$ with a duration longer than $d$. The reasons to consider
$P_m(d)$ instead of $p_s(d)$ are: (i) $P_m(d)$ gives curves smoother
than $p_m(d)$ does, making analyses easier.
(ii) $P_m(d)$ does not lose any information carried in $p_m(d)$, and
(iii) $P_m(d)$ preserve shapes for power-law and for exponential
functional forms.}

\subsection*{B. The coefficient of asymmetry}
The coefficient of symmetry of $A$ is defined as

\begin{equation*}
A =  \frac{1}{3} \biggl[\bigl(\frac{T_{WR} - T_{RW}}{T_{WR} + T_{RW}}\bigr)^2+ 
\bigl(\frac{T_{LW} - T_{WL}}{T_{LW} + T_{WL}}\bigr)^2 +
\bigl(\frac{T_{LR} - T_{RL}}{T_{LR} + T_{RL}}\bigr)^2\biggr]^{1/2},
\end{equation*}
where the $T_{mn}$ are elements in the transition probability matrix
defined in the Methods section.
For a completely symmetric matrix in which
$T_{mn}=T_{mn}$, $A=0$, while for a completely asymmetric matrix
in which one of $T_{mn}$ and $T_{nm}$ is equal to $0$, $A=1$.

\newpage


\section*{Acknowledgments}
We thank the NIH/National Center for Research Resource
(P41 RR 13622) for support. We also thank the SIESTA project
(funded by the European Commission DG XII, as
Biomed-2 project No. BMH4-CT97-2040 ``SIESTA'') for providing data.
We thank A. L. Goldberger and C.-K. Peng for helpful discussions
and comments in the manuscript. CCL thank J. Mullington for helpful
suggestions.

\newpage


\newpage


\begin{figure}
\centerline{\LARGE\bf FIGURE 1}
\vspace*{0.4cm}

\begin{center}
 \begin{minipage}[b]{0.7\linewidth}
 \epsfig{file=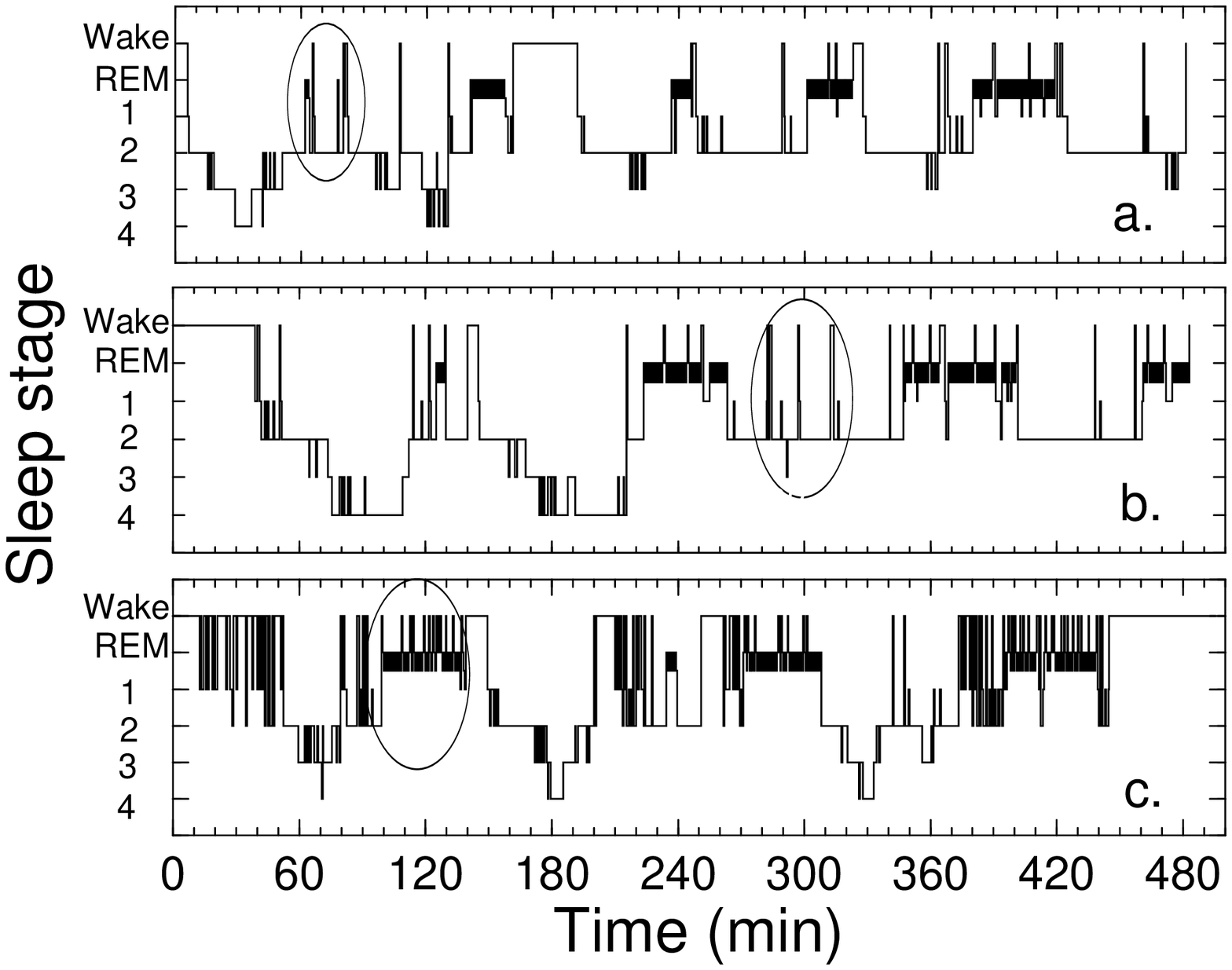, width=\linewidth}
\end{minipage}
\vspace{0.5cm}
\end{center}
\caption{Three typical hypnograms for normal subjects (a) and (b),
and for a sleep apnea subject (c). There are large number of short
sleep-stage transitions as shown in ovals throughout the nights
for both normal and sleep apnea subjects. The overall patterns of
hypnograms between different normal subjects are also very
different. The sleep apnea subject experience fragmented sleep,
and shows a much larger number of short transitions than normal
subjects do.} \label{f.hypnograms}
\end{figure}
\vfill
\eject

\begin{figure}
\centerline{\LARGE\bf FIGURE 2}
\vspace*{0.4cm}

\begin{center}
 \begin{minipage}[b]{0.65\linewidth}
 \epsfig{file=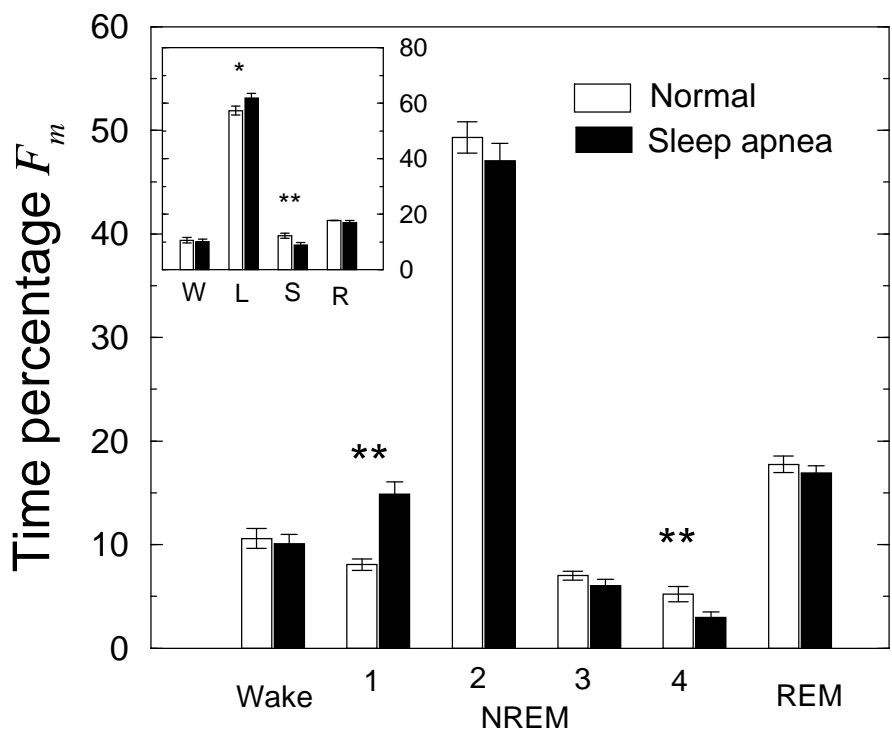, width=\linewidth}
\end{minipage}
\end{center}
\caption{Fraction $F_m$ of total sleep time for a given stage $m$.
Here we show the average values based on a database of 48 normal
subjects and 48 sleep apnea subjects with matched ages. The error
bars give uncertainties of the average values. Student's t-test is
performed to measure the significance of the difference between
normal and the sleep apnea groups. One asterisk indicates
$p<0.05$, and two asterisks indicate $p<0.01$. Sleep stages 1 and
4 display significant differences between normal and sleep apnea
subjects, while wake, stage 2, 3 and REM do not display
significant differences. Further on we consider only four stages:
wake (W), light sleep (L), slow-wave sleep (S), and REM sleep
(REM), therefore we show the percentage of total sleep time for
these four stages in the inset. } \label{f.fraction}
\end{figure}
\vfill
\eject

\begin{figure}
\centerline{\LARGE\bf FIGURE 3}
\vspace*{0.4cm}

\begin{center}
{\bf Normal}
 \begin{minipage}[b]{0.95\linewidth}
 \epsfig{file=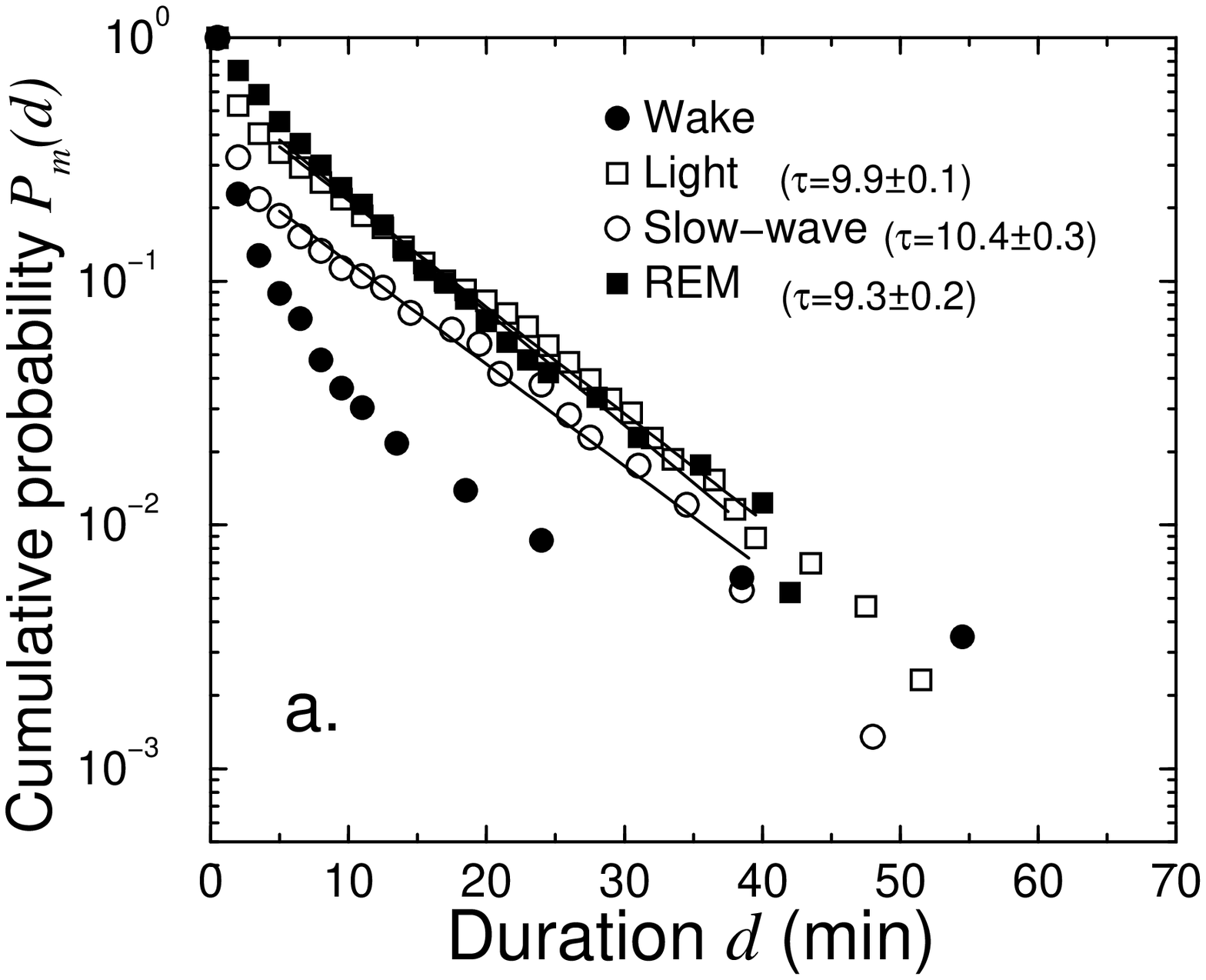, width=0.497\linewidth}
 \epsfig{file=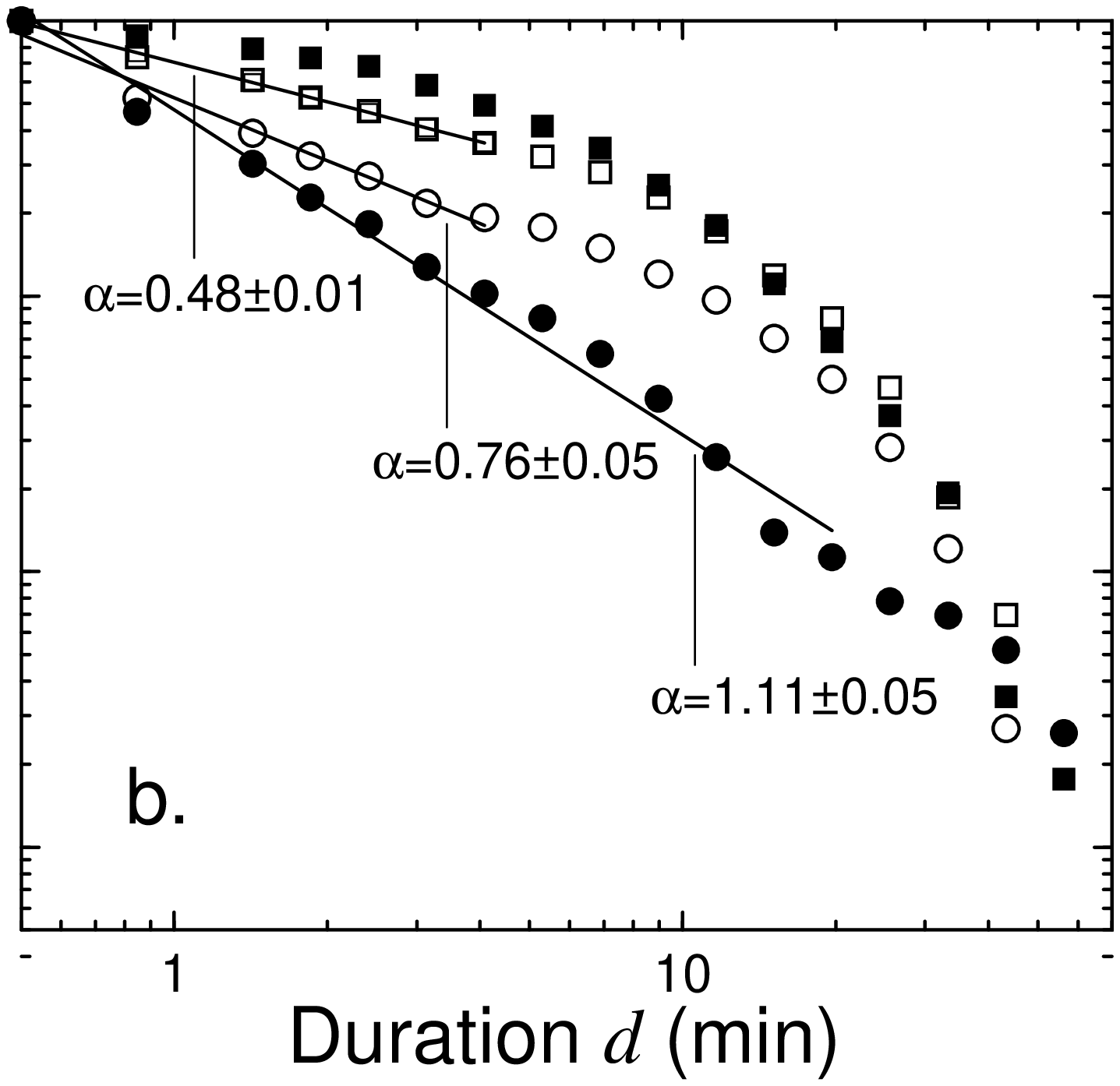, width=0.4\linewidth}
\end{minipage}\\
\vspace{1cm}
{\bf Sleep apnea}
 \begin{minipage}[c]{0.95\linewidth}
 \epsfig{file=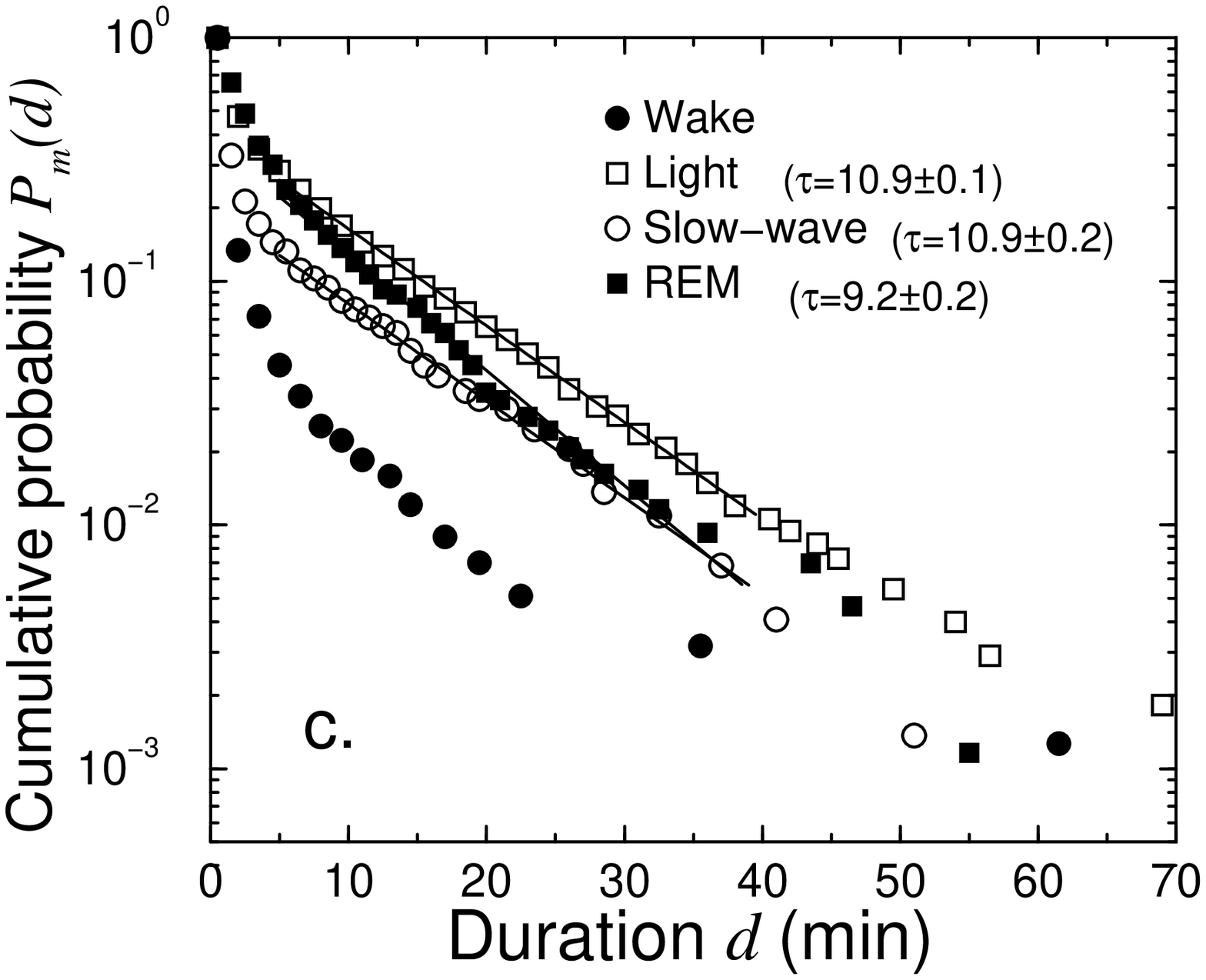, width=0.497\linewidth}
 \epsfig{file=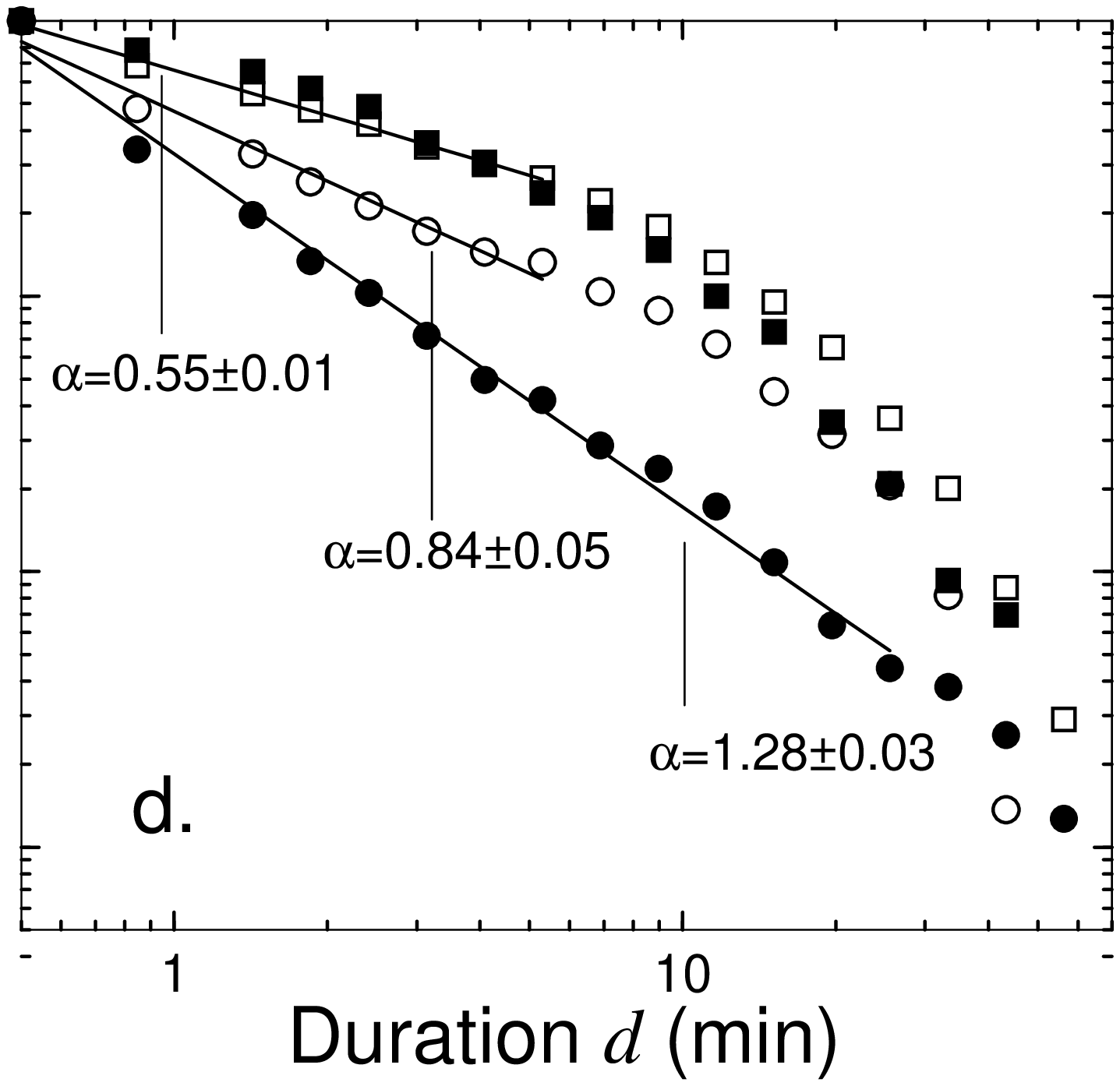, width=0.4\linewidth}
\end{minipage}

\end{center}
\caption{
Cumulative distribution of duration for each sleep
stage showing different features for wake stage and sleep
stages but similar features for normal and sleep apnea subjects.
(a) Semi-logarithmic plot and (b) double-logarithmic plot
show curves for the normal group. (c) Semi-logarithmic
plot and (d) double-logarithmic plot show curves for the
sleep apnea group. For both normal and sleep apnea groups,
the distributions for wake follow power-law
decays, while the distributions for all sleep stages
follow exponential decays. Comparing
to the normal group, the sleep apnea group shows larger power-law
exponent $\alpha$ for the distribution of wake duration, larger
characteristic time scale $\tau$ for the distribution of duration
of light sleep, but similar characteristic time scale for the
distributions of duration of slow-wave and REM sleep.
In order to compare curves of light, slow-wave and REM sleep in the
small time
region ($d<5$ min) between normal and sleep apnea subjects,
power-law functions are fit to the curves of light and slow-wave
sleep for $d<5$ min (c \& d). Note
that the fitting is only for the purpose of comparison.
The lack of data points in the region of $d<5$ min makes it
difficult to determine the functional form of the distribution
for sleep stages in this small-time region. Note that for
distributions of all sleep stages, the sleep apnea group shows
a steeper decay in the small time region ($d<5$ min).}
\label{f.distdur}
\end{figure}
\vfill
\eject

\begin{figure}
\centerline{\LARGE\bf FIGURE 4}
\vspace*{0.4cm}

\begin{center}
 \begin{minipage}[b]{0.95\linewidth}
 \epsfig{file=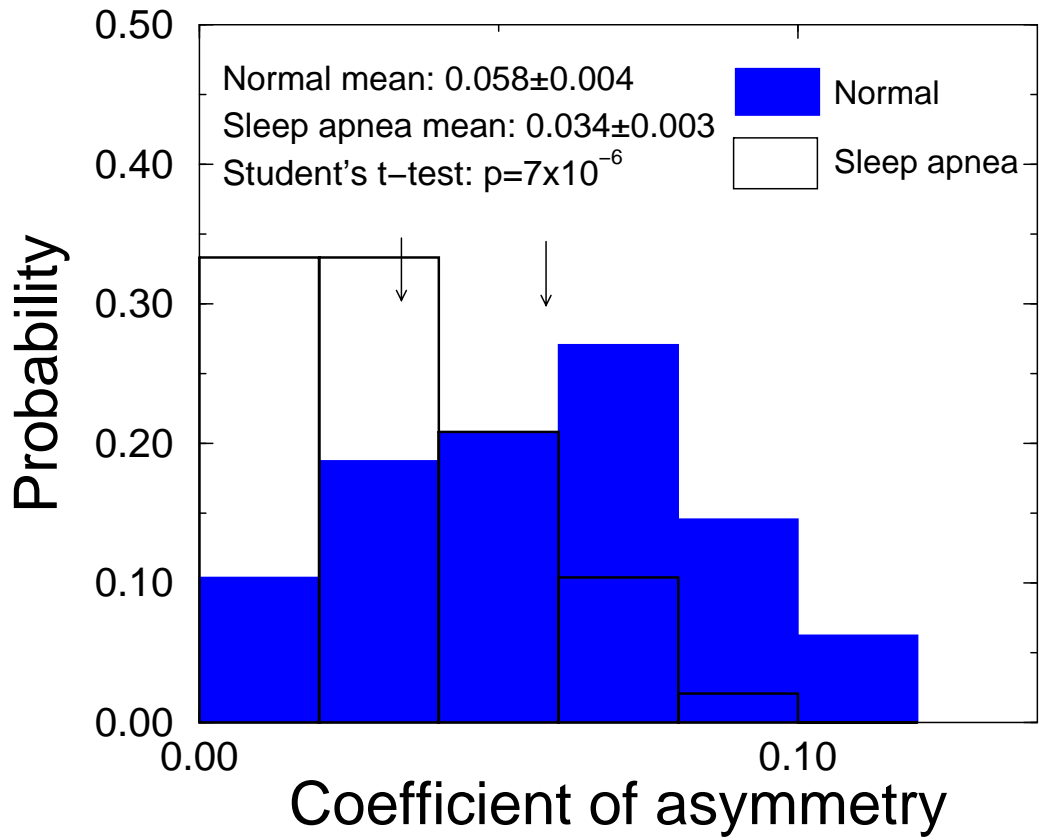, width=0.8\linewidth}
\end{minipage}

\end{center}
\caption{Distribution of the coefficient of symmetry
(defined in the text) for normal and sleep apnea groups.
Arrows indicate means of distributions of normal and sleep
apnea groups. The result of Student's t-test indicates
that the observed averages of these two groups are significantly
different, suggesting that the sleep apnea group has a significant
decrease of the asymmetry as the result of the sleep fragmentation.}
\label{f.asymmetry}
\end{figure}
\vfill
\eject

\begin{figure}
\centerline{\LARGE\bf FIGURE 5}
\vspace*{0.4cm}

\begin{center}
 \begin{minipage}[b]{0.95\linewidth}
\begin{center}
{\bf Normal}\\
 \epsfig{file=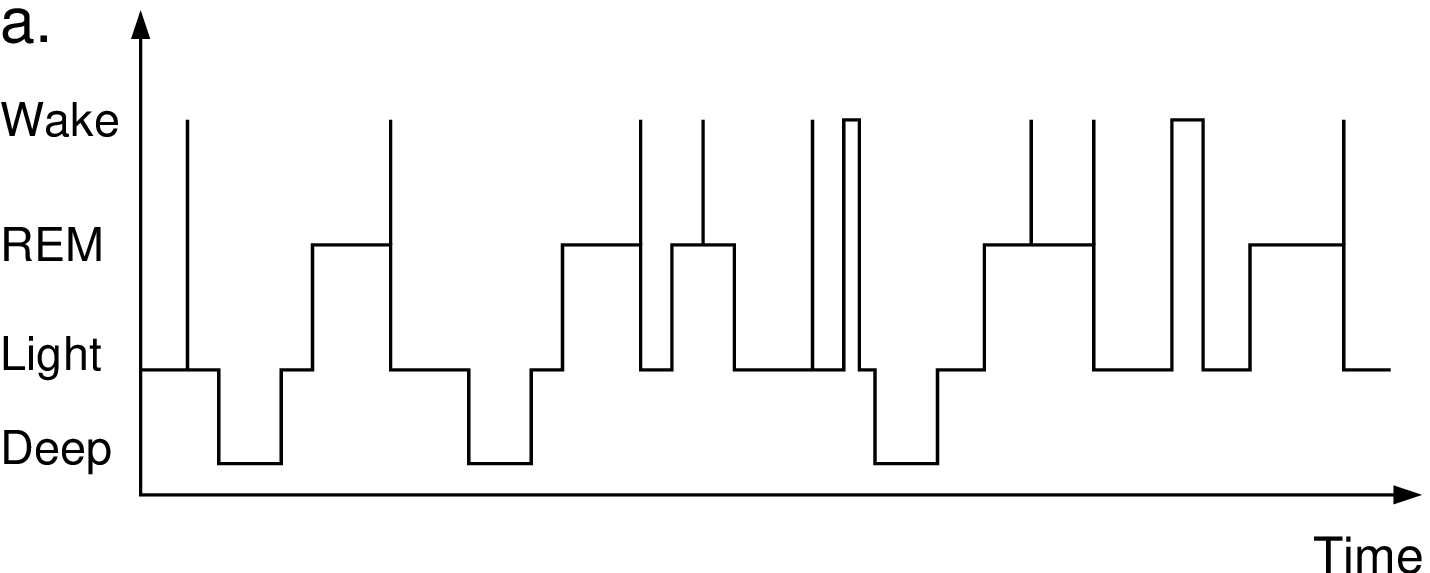, width=0.7\linewidth}
 \vspace{1.5cm}\\
{\bf Sleep Apnea}\\
 \epsfig{file=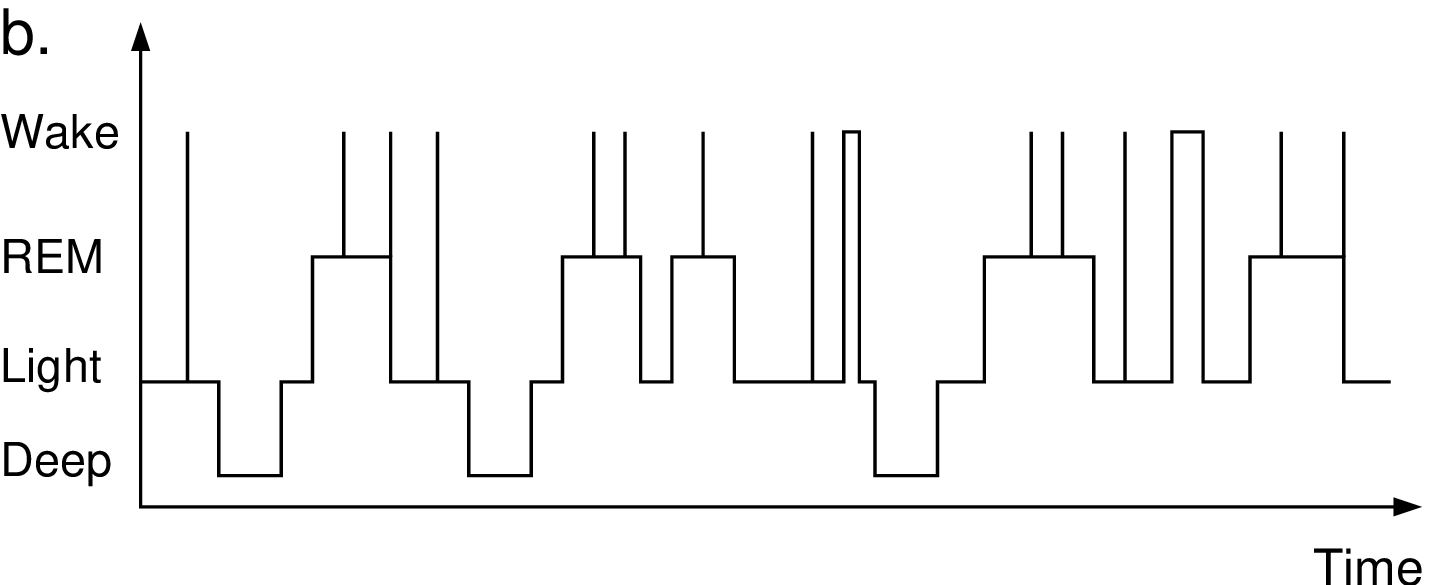, width=0.7\linewidth}
\end{center}
\end{minipage}
\end{center}
\caption{Illustration of asymmetric sleep-stage transitions.
Since wake periods are much shorter than the light,
slow-wave and REM sleep periods on average,
we assume that the basic structure of sleep-stage transitions
are dominated by light sleep, slow-wave sleep and REM sleep.
(a) Wake periods can be viewed as spikes distributed
throughout the night in light and REM sleep. Because
$T_{WR} > T_{RW}$ for normal subjects (cf. Table 2, where
$T_{WR}/T_{RW}\approx 5.6$), there is a preference for
transitions from REM to wake and then to light sleep, instead
of back to REM. (b) For sleep apnea subjects, there is increasing
symmetry in transitions (cf. $T_{WR}/T_{RW}\approx 2.6$). Because
sleep apnea subjects have more transitions on average, the
increased wake periods may distribute in light and REM
sleep with less preference throughout the night.
}
\label{f.illust}
\end{figure}
\vfill
\eject

\centerline{\LARGE\bf Table 1} \vspace*{0.4cm} a) Transition
probability matrix $T_{mn}$ (defined in the text) for normal subjects,
where $m$ corresponds to row and $n$ corresponds to column.
The numbers
in the matrix are means of the group distributions and
standard errors of means.
\\ The average number of transitions per night  $= 96.0 \pm 3.2$.
\begin{center}
\begin{gather*}
\begin{matrix}
  &  & W & & R  & & L  & & S \\ & & & & & & & & \\
W & & - & & 0.050\pm 0.005 & & 0.182\pm 0.009 & & 0.016\pm 0.002\\
R & & 0.009\pm 0.002 & & - & & 0.116\pm 0.007 & & 0.004\pm 0.001 &\\
L & & 0.237\pm 0.010 & & 0.073\pm 0.006 & & - & & 0.139\pm 0.010\\
S & & 0.001\pm 0.001 & & 0.000\pm 0.000 & & 0.155\pm 0.010 & & -\\
\end{matrix}
\label{t.1a}
\end{gather*}
\end{center}

b) Same as above for sleep apnea.\\
The average number of transitions per night $= 123.0 \pm 6.2$.
\begin{center}
\begin{gather*}
\begin{matrix}
 &  & W & & R & & L & & S \\
 &  &   & &   & &   & &  \\
W &  & - & & 0.042\pm 0.004 & & 0.217\pm 0.014 & & 0.008\pm 0.002\\
R &  & 0.016\pm 0.004 & & - & & 0.126\pm 0.012 & & 0.001\pm 0.001\\
L &  & 0.250\pm 0.014 & & 0.099\pm 0.011 & & - & & 0.105\pm 0.010\\
S &  & 0.001\pm 0.001 & & 0.000\pm 0.000 & & 0.114\pm 0.010 & & -\\
\end{matrix}
\label{t.1b}
\end{gather*}
\end{center}

\eject

\centerline{\LARGE\bf Table 2}
{\noindent Table 2. Summary of results of our analysis for
(i) the mean time percentage, (ii) the distribution of duration
of sleep stages and (iii) the mean degree of asymmetry of the
transition probability matrix.}

\begin{tabular}{c c c c c c c c c c c }
\multicolumn{10}{l} {}\\\hline
&\multicolumn{4}{c}{$\bar{F}_m$ (\%)} & { }
&\multicolumn{4}{c}{$P_m(d)$}&
\\ \cline{2-5} \cline{7-10}
\raisebox{1.5ex}[0cm][0cm]{Subjects}
& W & L* & S* & R & & W* & L* & S & R &\raisebox{1.5ex}[0cm][0cm]{$\bar{A}$*} \\
 \hline
Normal & \hspace{0mm} 10.6 \hspace{0mm} & \hspace{0mm} 57.3  \hspace{0mm}
 & \hspace{0mm} 12.2  \hspace{0mm} &
 \hspace{0mm} 17.8  \hspace{0mm} & \hspace{0mm}
 &  \hspace{0mm} $d^{-1.1}$  \hspace{0mm} &  \hspace{0mm} $e^{-d/9.9}$ \hspace{0mm}
 &  \hspace{0mm} $e^{-d/10.4}$ \hspace{0mm}  &  \hspace{0mm} $e^{-d/9.3}$
 \hspace{0mm}& 0.58 \\
Sleep Apnea & 10.1 & 61.8 & 8.9 & 17.0 & & $d^{-1.3}$ & $e^{-d/10.9}$
 & $e^{-d/10.9}$ & $e^{-d/9.2}$ & 0.34 \\\hline
\end{tabular}

\vspace{0.5cm}
{\noindent
Symbols: $\bar{F}_m$, mean time percentage for stage $m$. $P_m(d)$,
distribution of duration $d$ of stage $m$. $\bar{A}$, mean degree of
asymmetry. Stage $m$ can be wake ($W$), light
sleep ($L$), slow-wave sleep ($S$) or REM ($R$).
}

{\noindent
An asterisk denotes significant difference between normal and sleep apnea groups.
}
\vfill
\eject

\end{document}